\documentclass[apj]{emulateapj}
\submitted{Accepted for publication in The Astrophysical Journal}
\def\ergs{erg~s$^{-1}$}
\def\ergcms{erg$~$~cm$^{-2}$~s$^{-1}$}
\usepackage{color} 
\begin{document}

\title{XMM-Newton Observations of NGC 247: X-ray Population and a Supersoft Ultraluminous X-ray Source}

\author{Jing Jin\altaffilmark{1}, Hua Feng\altaffilmark{1}, Philip Kaaret\altaffilmark{2}, Shuang-Nan Zhang\altaffilmark{3}}
\altaffiltext{1}{Department of Engineering Physics and Center for Astrophysics, Tsinghua University, Beijing 100084, China}
\altaffiltext{2}{Department of Physics and Astronomy, University of Iowa, Van Allen Hall, Iowa City, IA 52242, USA}
\altaffiltext{3}{Key Laboratory of Particle Astrophysics, Institute of High Energy Physics, Chinese Academy of Sciences, Beijing 100049, China}

\shortauthors{Jin et al.}
\shorttitle{XMM-Newton Observations of NGC 247}

\begin{abstract}
We report on a new XMM-Newton observation of NGC 247 from December 2009. The galaxy contains a supersoft, ultraluminous X-ray source (ULX) whose spectrum consists of a thermal component with a temperature about 0.1~keV and a power-law tail with a photon index around 2.5. The thermal emission is absolutely the dominant component, contributing 96\% of the total luminosity in the 0.3-10 keV band. Variability is detected at timescales of $10^2$~s and longer with a $\nu^{-1}$ power spectrum. These properties are consistent with black hole binaries in the thermal state and suggest the presence of an intermediate mass black hole of at least 600 solar masses. However, the integrated rms power is much higher than typically found in the thermal state. An alternative explanation of the emission could be a photosphere with a radius about $10^9$~cm. A possible absorption feature around 1~keV is detected, which may be due to absorption of highly ionized winds. X-ray sources within the disk of NGC 247 have a luminosity function consistent with that found in low mass X-ray binaries. We confirm previous results that X-rays from the quasar PHL~6625 may be absorbed by gas in NGC 247, mainly at energies below 0.3~keV.

\end{abstract}


\section{Introduction}

Nonnuclear X-ray sources with luminosities higher than $3 \times 10^{39}$~\ergs\ are classified as ultraluminous X-ray sources (ULXs). Many of them are variable and most likely powered by accretion onto black holes. However, their luminosities, assuming isotropic emission, are higher than the maximum seen in Galactic black hole binaries, suggesting the presence of intermediate mass black holes with masses of $10^2 - 10^4$~$M_\sun$ \citep{col99,kaa01,far09} or super-Eddington accretion onto stellar mass black holes of about 10~$M_\sun$ \citep{wat01,beg02}. Discovery of intermediate mass black holes would be important because they cannot be formed via core collapse of a single star with normal metallicity \citep{bel10} and may shed light onto the formation of supermassive black holes in the early Universe \citep{ebi01,vol10}.

Dynamical measurement of the compact object mass for ULXs has been unfeasible to date \citep{rob10}. Instead, indirect means such as X-ray spectroscopy may shed light on the black hole mass. At a fixed Eddington ratio, the inner temperature of the accretion disk scales with black hole mass as $T_{\rm in} \propto M^{-1/4}$ \citep{mak00}. Application of this relation requires the black hole binary to be in the thermal state \citep{mcc06,rem06}, in which the disk is believed to extend all the way to the last stable orbit around the black hole. However, the thermal state is rarely found in ULXs, the only known occurrences are M82 X41.4+60 \citep{fen10} and M82 X37.8+54 \citep{jin10} during outburst.  X-ray sources with soft, $kT_{\rm in} \sim 0.1$~keV, thermal spectra may be good candidates for intermediate, $\sim 10^3 \, M_\sun$, mass black holes in the thermal state.

NGC 247 is a nearby spiral galaxy in the Sculptor Group, with a modest star formation rate estimated to be about 0.1~$M_\sun$~yr$^{-1}$ by counting bright main-sequence stars \citep{dav06} or based on H$\alpha$ observations \citep{fer96}. However, infrared and radio observations suggest a star formation rate 10 or 100 times lower \citep{dav06}. The distance to NGC 247 has been measured independently by different groups using various means. Among them, the recent result from \citet{gie09} using near-infrared observations of Cepheids is claimed to be the most accurate. We therefore adopt that distance of 3.4~Mpc in this paper.

Previously, NGC 247 has been observed in the X-ray band with Einstein \citep{fab92}, ROSAT \citep{zan97,rea97,lir00}, and XMM-Newton \citep{win06}. A bright, off-nucleus X-ray source (NGC 247 ULX = 1RXS~J004704.8$-$204743) was detected in all of these observations. ROSAT observations revealed a very soft spectrum with $kT = 0.12$~keV \citep{rea97}.  Later on, the source was observed to have brightened by a factor of 2 \citep{lir00}, suggestive of an accretion powered system. In 2001, XMM-Newton detected the source at the highest flux observed to date, with an unabsorbed luminosity of $8 \times 10^{39}$~\ergs\ (at 3.4~Mpc) in the 0.3-10 keV band, and the spectrum was still dominated by a soft thermal component with $kT = 0.12$~keV \citep{win06}.

These results imply that the source is a good candidate to be an intermediate mass black hole in the thermal state.  Unfortunately, the 2001 XMM-Newton observation suffered strong background flares, resulting in little usable data and poor statistics in the extracted spectrum. Thus, we proposed a new XMM-Newton observation of NGC 247 to obtain a better spectral measurement. In this paper, we report results from the new observation about the ULX as well as other sources in NGC 247.

\section{Observations and results}

The observation was performed on 2009 Dec 27 (PI: H.~Feng), lasting about 35~ks, with the prime full window for the three cameras. The thin filter was chosen for the PN and medium filter for the MOS. Event lists were created from the observation data files using SAS with recent calibrations. To minimize background contamination, we selected good time intervals (GTIs) where the 10-15 keV count rate of the whole CCD was lower than the $3\sigma$ upper limit of the quiescent level.  This resulted in 22~ks of clean exposure for the PN and 30~ks for the MOS.

\begin{figure*}
\centering
\includegraphics[width=0.7\textwidth]{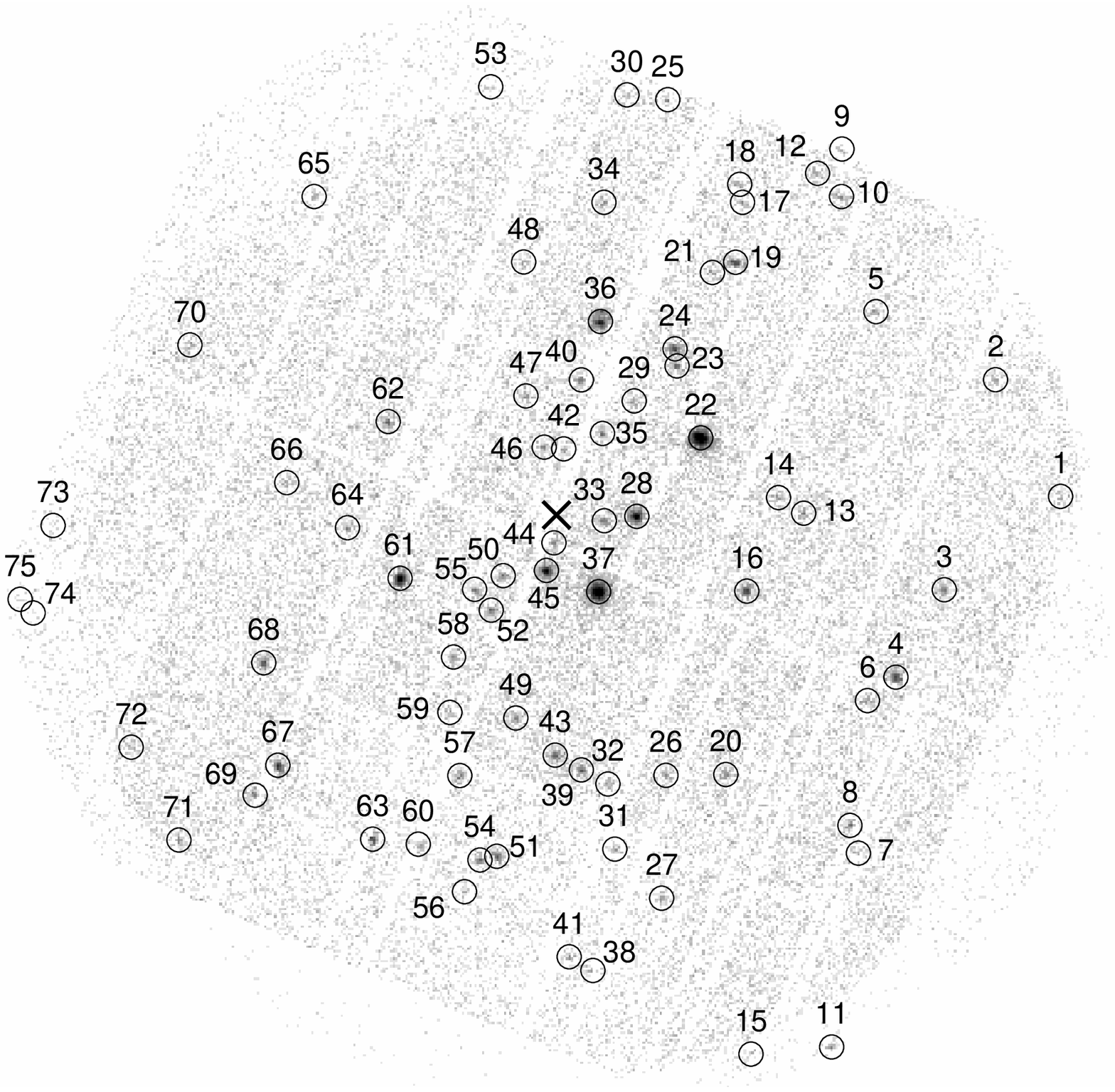}\\
\includegraphics[width=0.7\textwidth]{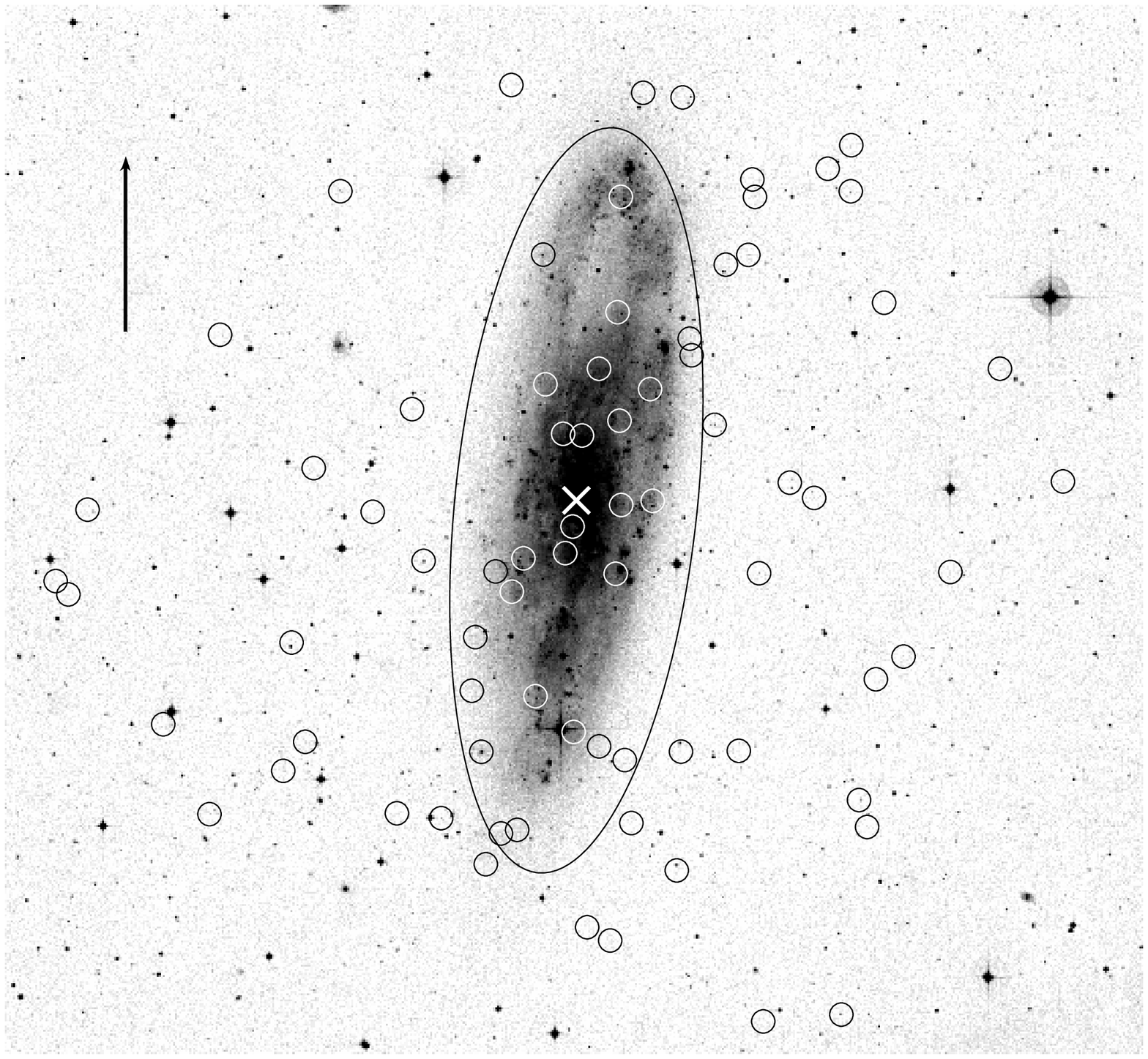}
\caption{X-ray (left) and optical (right) images of NGC 247. The X-ray image is created from the combined PN and MOS data and circles indicate detected sources. Source 22 is PHL 6625 and source 37 is the ULX. The cross indicates the galaxy center. The optical image is from the Digitized Sky Survey blue band and the ellipse indicates the D25 region of NGC 247. The arrow points north and has a length of 5\arcmin.
\label{fig:img}}
\end{figure*}

Images in two energy bands, 0.3-2~keV and 2-10~keV, were created for each CCD with a pixel size of 4.1\arcsec\ from events with FLAG = 0 and PATTERN $\leq$ 4 for the PN or PATTERN $\leq$ 12 for the MOS. The {\tt edetect\_chain} tool was used to detect point sources simultaneously on the six images. A total number of 75 sources with likelihood (=$-\ln(p)$, where $p$ is the chance probability) over 10 were detected and are listed in Table~\ref{tab:src} and shown in Figure~\ref{fig:img}. The astrometry was corrected by aligning the X-ray position of the known quasi-stellar object (QSO) PHL~6625 (source 22) to its optical position quoted in the USNO-B1.0 catalog.

The absorbed flux of each source was calculated from the net counts and an effective exposure with a response matrix created at the source position on the CCD assuming a power-law spectrum with Galactic absorption \citep[$N_{\rm H} = 2.01 \times 10^{20}$~cm$^{-2}$;][]{kal05} and a photon index of 1.7. We note that this is more accurate than using a single conversion factor in {\tt edetect\_chain} as the energy band is wide. In the calculation, either the PN or the averaged MOS count rate was adopted depending on which likelihood was higher. For sources 38, 60, and 63, only the MOS2 data was adopted as they were not covered by the MOS1. The hardness ratio was calculated using {\tt edetect\_chain} for each source as $(R_2 - R_1) / (R_2 + R_1)$, where $R_1$ and $R_2$ are vignetting-corrected net count rates in the soft and hard bands, respectively.

For 15 sources, optical or infrared sources in the USNO-B1.0 or Two Micron All Sky Survey catalogs are found in a 2\arcsec-radius circle around their corrected X-ray positions, and are indicated in Table~\ref{tab:src}. Most of the sources with optical/IR counterparts, 11 out of 15, are outside of the D25 ellipse of NGC 247.

For sources with 200 or more net counts, we extracted the PN and the MOS spectra and fitted them simultaneously with a power-law model subject to interstellar absorption. The lower limit of the absorption column density was set to the Galactic value. For sources 4 and 67, only the MOS spectra were adopted as they lie on a gap or bad pixels on the PN. For PHL~6625, only the PN data were adopted for the same reason, and a redshifted power-law model with $z = 0.38$ was used.

The best-fit spectral parameters are listed in Table~\ref{tab:top}, except for the ULX. Among them, seven sources (4, 16, 19, 22, 39, 61, and 62) are coincident with an optical counterpart in USNO-B1.0, and source 67 has two possible counterparts. We calculated their X-ray to optical flux ratios as $\log(f_{\rm X}/f_V) = \log f_{\rm X} + m_V/2.5 + 5.37$, where $f_{\rm X}$ is the observed flux in the 0.3-3.5 keV band in units of \ergcms\ and $m_V$ is the visual magnitude \citep{mac82}. The average of the second red and blue magnitudes in USNO-B1.0, if available, was used to estimate the visual magnitude, otherwise, the red or blue magnitude is taken as an approximation to $m_V$. The X-ray to optical flux ratios are 0.52, 0.05, 0.24, 0.33, $-0.28$, $-0.26$, 0.04, respectively for sources 4, 16, 19, 22 (PHL~6625), 39, 61, 62.  For source 67, the ratio is 0.20 or 0.54 depending on the counterpart. These values are consistent with those found for active galactic nuclei (AGN) and clusters of galaxies \citep{mac88,sto91}; the ratios for sources 39 and 61 are also consistent with those of normal galaxies, and the ratios for source 4 and source 67 (with the dimmer optical counterpart) are also consistent with those of BL Lac objects.

For other sources that are associated with an optical counterpart but do not have enough photons for spectral fitting, we calculated their 0.3-3.5 keV fluxes from count rates and the X-ray to optical flux ratios were found in a range from $-2.2$ to $-0.2$, which are consistent with those of AGN or normal galaxies, but smaller than those of X-ray binaries. Two of them (source 48 and source 57 with the brighter optical counterpart, both in D25) with ratios smaller than $-1.5$ could also be foreground M or K dwarfs in the halo of the Milky Way. Thus, we conclude that sources associated with bright optical sources are unlikely to be members of NGC 247, and more likely to be background or foreground objects.

\begin{figure}
\centering
\includegraphics[width=\columnwidth]{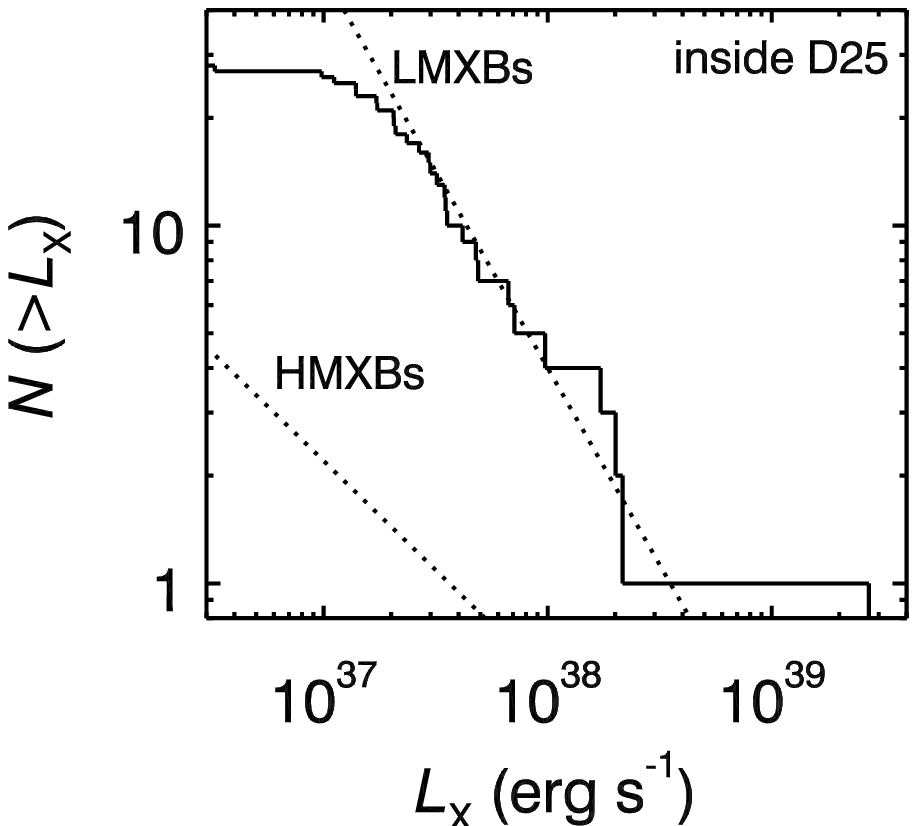}
\caption{Cumulative X-ray luminosity functions of NGC 247 for sources inside the D25 ellipse. The steep dotted line is the luminosity function for LMXBs 
with a slope of $-1.1$  normalized to the observed number at $10^{38}$~\ergs.  The shallower dotted line is the luminosity function for HXMBs with a slope of $-0.61$ normalized to a star formation rate of 0.1~$M_\sun$~yr$^{-1}$.
\label{fig:lum}}
\end{figure}

The cumulative X-ray luminosity function (XLF) for sources inside the D25 ellipse of NGC 247 is plotted in Figure~\ref{fig:lum}. For sources with enough photons, those in Table~\ref{tab:top} and the ULX, the flux in the 0.3-10 keV band from spectral fitting is adopted, otherwise, the two fluxes in Table~\ref{tab:src} after correction for Galactic absorption are summed and adopted. For comparison, two model XLFs are shown. The shallower curve is the XLF expected for high mass X-ray binaries (HMXBs): a power-law with a slope of $-0.61$ as found for HMXBs in nearby starburst galaxies normalized to a star formation rate of 0.1~$M_\sun$~yr$^{-1}$ \citep{gri03,dav06,fer96}. The steeper curve is the XLF for low mass X-ray binaries (LMXBs), a power-law with a slope of $-1.1$ \citep{kim04}, normalized to the observed number of sources at $L_{\rm X} > 10^{38}$~\ergs.  Comparing the curves with the data, it is clear that the X-ray binary population in NGC 247 is dominated by LMXBs.

\begin{figure}
\centering
\includegraphics[width=\columnwidth]{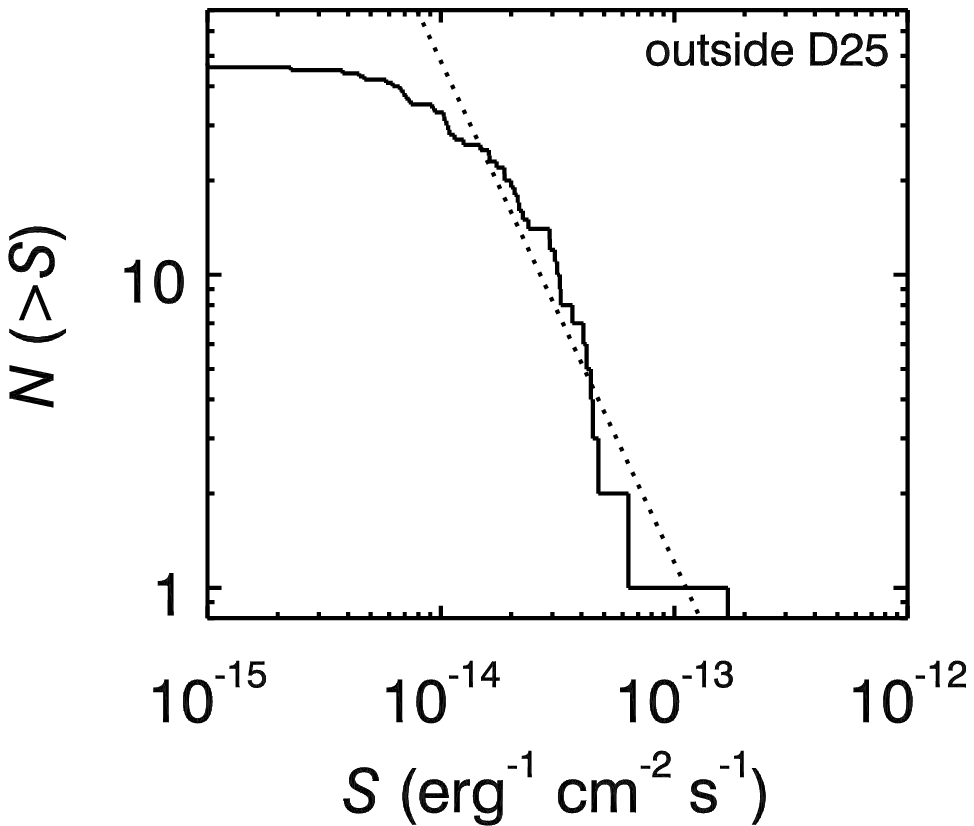}
\caption{Number of sources, $N (>S)$, brighter than a given observed flux in the 2-8 keV band, $S$, for sources outside of the D25 ellipse of NGC 247. The dotted line is the distribution for AGN with a slope of $-1.6$ normalized to the observed solid angle of about 0.16~degree$^2$.
\label{fig:logns}}
\end{figure}

For sources outside of the D25 ellipse, we plotted $\log N (>S)$ versus $\log S$, where $S$ is the observed flux in the 2-8 keV band and $N(>S)$ is the number of sources with flux greater than $S$, see Figure~\ref{fig:logns}. For comparison, we show a power-law index of $-1.6$ as expected for AGN \citep{bra05}.  The power-law is normalized to a solid angle of 0.16~degree$^2$, equal to the sky area outside the D25 ellipse of NGC 247 and in the field of view of XMM-Newton.

We examined short-term variability for each source using photons in the 0.3-10 keV band with the Kolmogorov-Smirnov test. Only the ULX shows evidence for variability with a chance probability lower than $10^{-3}$.

\subsection{Spectral and timing analysis of the ULX}

Due to the soft spectrum of the ULX and the thin filter on the PN, the number of photons detected on the PN is about 3-4 times that on each MOS. Thus, we used only the PN data for spectral analysis. The energy spectrum was extracted from events in a 30\arcsec-radius circular region around the source to avoid CCD gaps and with FLAG and PATTERN the same as those used in creating the image. Background was subtracted from events in a nearby circular region with a radius of 45\arcsec\ on the same CCD and off the readout column of the ULX. To avoid spurious spectral structure caused by over-sampling of the spectrum, the energy channels were grouped with a bin size of 1/5th of the local spectral resolution (in terms of full width at half maximum) and to have at least 15 net counts per bin.

\begin{figure}
\centering
\includegraphics[width=\columnwidth]{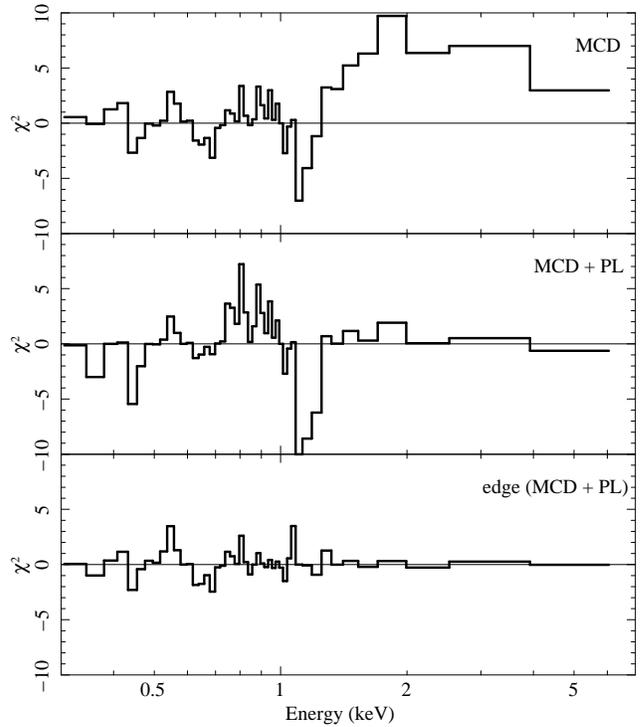}
\caption{Residuals in units of $\chi^2$ from fits to the spectrum of NGC 247 ULX with models
wabs $\ast$ diskbb, wabs $\ast$ (diskbb + powerlaw), wabs $\ast$ edge $\ast$ (diskbb + powerlaw), respectively.
\label{fig:res}}
\end{figure}

A single power-law subject to interstellar absorption is inadequate to fit the ULX spectrum, resulting in $\chi^2 = 198.0$ for 44 degrees of freedom (dof). An absorbed multicolor disk (MCD) model provides a better fit with $\chi^2 / {\rm dof} = 98.5/44$, but is still inadequate, especially at energies above 1.4~keV where a hard excess is obvious, see the top panel in Figure~\ref{fig:res}. Therefore, we used a model with the sum of a power-law and a MCD, which significantly improved the fit with $\chi^2 / {\rm dof} = 63.4/42$. To test the significance of the power-law tail, we simulated $10^4$ spectra based on the single MCD model.  In none of the simulated spectra did addition of a power-law reduce the $\chi^2$ by the amount seen in the real data.  This suggests a chance probability less than $10^{-4}$. The F-test gives a consistent probability of $9.5 \times 10^{-5}$ despite caveats about its validity in such a case \citep{pro02}.

Although the MCD plus power-law model improves the fit considerably, it still does not provide an adequate fit. As one can see from the middle panel of Figure~\ref{fig:res}, most residuals arise around 1~keV and appear like an absorption feature. Thus, we then tried two absorption components, both of which gave adequate fits.  An absorption edge at 1.02~keV gave $\chi^2 / {\rm dof} = 36.8/40$ and a Gaussian absorption line with a centroid of 1.16~keV gave $\chi^2 / {\rm dof} = 37.0/39$. The residuals for both models are flat throughout the passband, see the bottom panel of Figure~\ref{fig:res}. The absorption equivalent width is 0.13 to 0.19~keV, depending on the model. Similarly, we tested the significance of the absorption component using simulations, and an edge was detected as significant in only 1 out of $10^4$ trails, indicative of a chance probability of $\sim 10^{-4}$.

\begin{figure}
\centering
\includegraphics[width=\columnwidth]{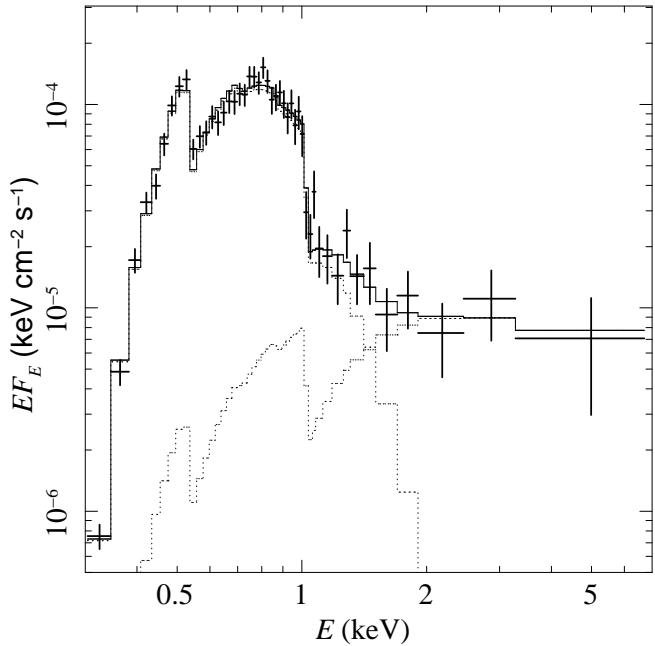}
\caption{The unfolded spectrum of NGC 247 ULX fitted with the MCD plus power-law model with edge and interstellar absorption.
\label{fig:unf}}
\end{figure}

The spectral parameters are listed in Table~\ref{tab:spe}. Due to the poor fit of the MCD plus power-law model, we fixed the absorption column density at the best value, $N_{\rm H} = 3.2 \times 10^{21}$~cm$^{-2}$, obtained from the best-fit model. Consistent parameters of the continuum components were obtained for all models. The unfolded spectrum from the MCD plus power-law model with absorption edge is plotted in Figure~\ref{fig:unf}. The MCD is the dominant component of the spectrum, with a contribution of 96\% of the total unabsorbed flux in the 0.3-10 keV band.

We also replaced the MCD with a blackbody and obtained similar results as shown in Table~\ref{tab:spe}. We note that the use of a more sophisticated model for interstellar absorption, such as {\tt tbabs}, leads to almost identical results as obtained with {\tt wabs} due to the moderate resolution and statistics.

We tried to fit the spectrum with a MCD plus Comptonization model, which is found to fit for ULX spectra with high statistics \citep{gla09}. 
This model provides an equally good fit as the MCD plus power-law model and results in almost identical parameters for the disk component. The disk temperature is $0.13\pm0.02$ keV and the inner disk radius is
$0.9_{-0.5}^{+0.9}\times10^4$ km assuming a face-on orientation.
A similar edge is required in the model at $1.02_{-0.03}^{+0.02}$ keV, with optical depth of $1.4_{-0.2}^{+0.5}$.
The Comptonization parameters are poorly constrained due to the low number of photons in the 2-10 keV band.

\begin{figure}
\centering
\includegraphics[width=\columnwidth]{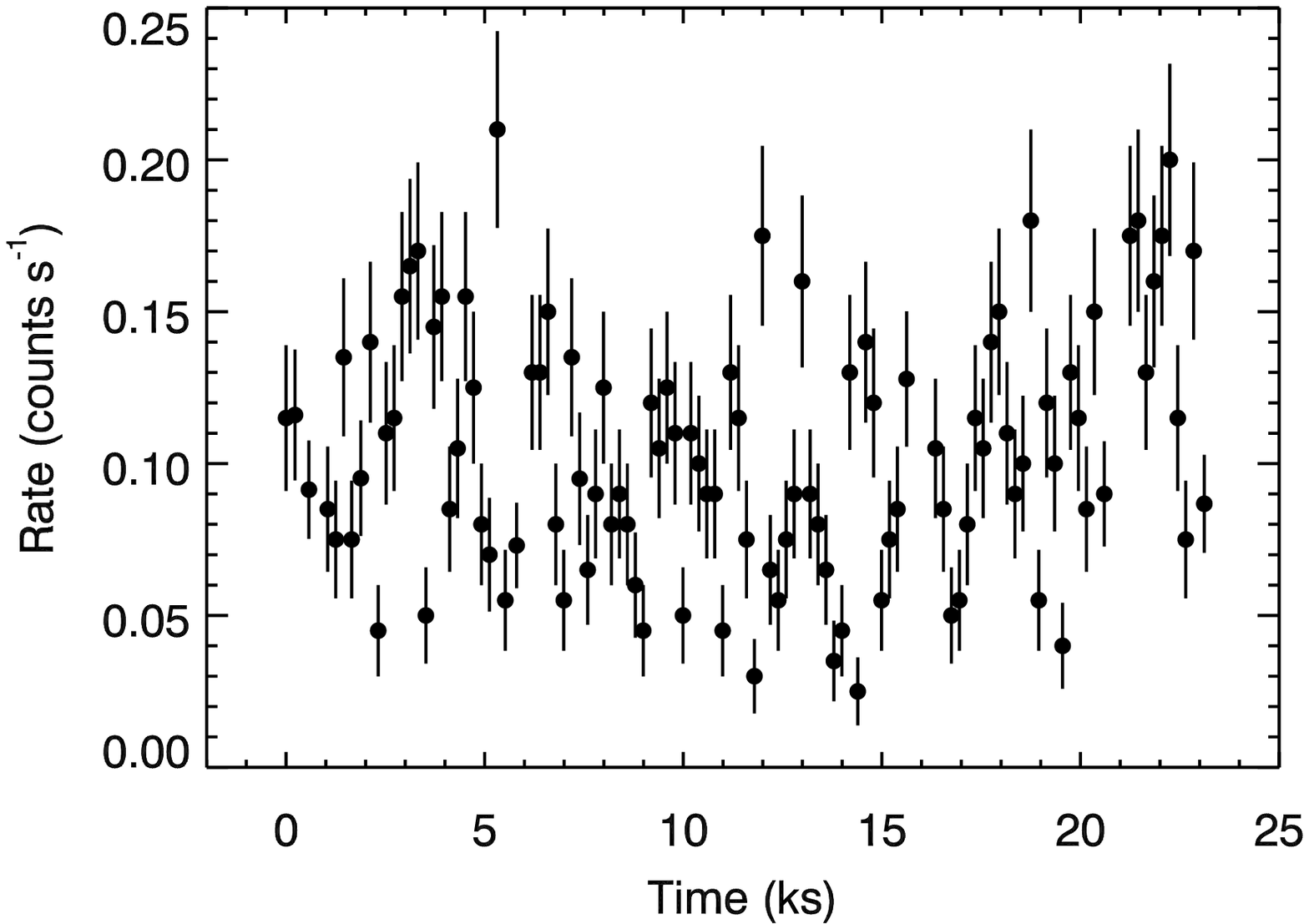}
\includegraphics[width=\columnwidth]{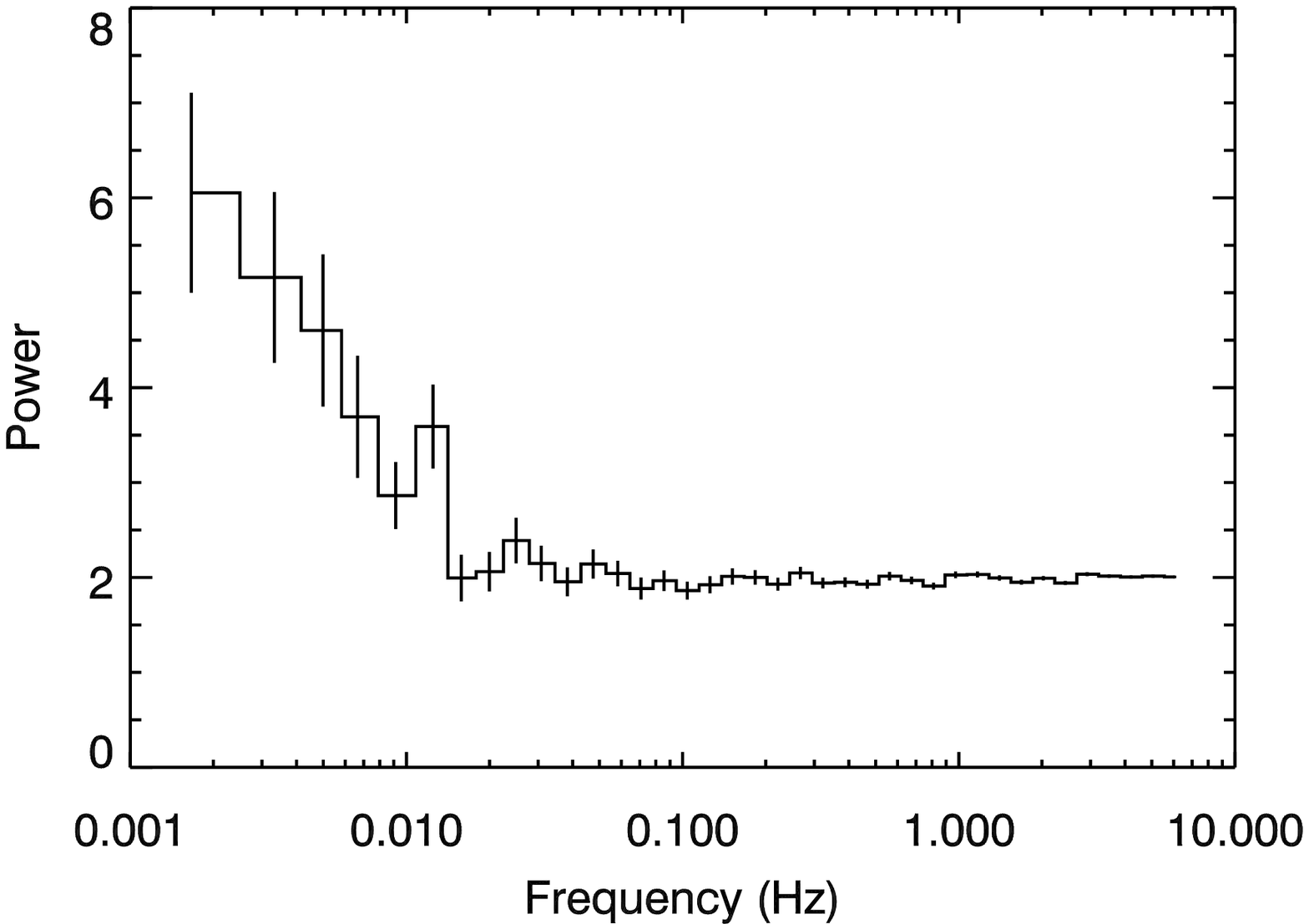}
\caption{X-ray lightcurve and power spectrum of NGC 247 ULX in the energy range of 0.3-2 keV. The lightcurve has a time bin size of 200~s.
\label{fig:pow}}
\end{figure}

Light curves of the ULX were created using the PN events in the same region as for spectral analysis with FLAG = \#XMMEA\_EP and PATTERN $\le$ 4.  Continuous intervals with low background were selected from within the GTIs. The 0.3-2~keV light curve shown in Figure~\ref{fig:pow} has a bin size of 200~s.

For each continuous interval, power spectra were calculated from a light curve with a bin size equal to the CCD frame time, $\sim$73~ms, using 8192-point Fourier transforms.  A final power spectrum was created by averaging all the individual power spectra with frequency channels geometrically binned by a factor of 1.2. Each point in the final spectrum is the mean of at least 33 individual data points, to assure validity in modeling the spectrum using minimal $\chi^2$. The final power spectrum in the 0.3-2 keV band normalized following \citet{lea83} is shown in Figure~\ref{fig:pow}. A power-law plus a constant can adequately fit the power spectrum, with the constant describing the Poisson noise level of $1.995 \pm 0.009$ and the power-law describing low-frequency noise with a slope of $1.22_{-0.16}^{+0.20}$. The source count rate is 0.10 counts~s$^{-1}$ in the 0.3-2 keV band and the background photon fraction is estimated to be 4.1\%. The fractional root-mean-square (rms) power of the source is estimated to be 0.52 by integrating the power-law component over 0.001-0.1~Hz, or 0.31 extrapolating the power-law to the frequency range 0.1-10~Hz. The power spectrum in the 2-10 keV band is consistent with that of Poisson noise, due to poor statistics.

We re-analyzed the XMM-Newton observation of NGC 247 obtained in 2001 in a similar way. As the background is too high to find a quiescent level adaptively for the PN data, an additional requirement that the background count rate is less than 1.5~counts~s$^{-1}$ for the PN is imposed to generate GTIs. The PN and the MOS spectra are fitted simultaneously with a MCD plus power-law model, resulting in $\chi^2 / {\rm dof} = 53.4/49$. The MCD component contributes about 88\% of the unabsorbed luminosity in the 0.3-10 keV band, which is $L_{\rm X} = 8_{-4}^{+13} \times 10^{39}$~\ergs, with a disk inner temperature of $0.15 \pm 0.03$~keV and a face-on inner radius of $1.2_{-0.7}^{+2.5} \times 10^4$~km. The photon index of the power-law component is poorly constrained, with a 90\% upper limit of 3.8. The interstellar absorption has a column density of $3.6_{-1.0}^{+1.3} \times 10^{21}$~cm$^{-2}$. The best-fit parameters are consistent with those reported in \citet{win06} within errors.

\section{Discussion and conclusion}

\subsection{X-ray population in NGC 247}

The XMM-Newton observation of NGC 247 obtained in December 2009 is, to date, the deepest X-ray observation of the galaxy and allows an X-ray source population study.  As shown in Figure~\ref{fig:lum}, sources inside the D25 ellipse of NGC 247 have a luminosity function well consistent with that of LMXBs. A population of X-ray sources dominated by LMXBs is reasonable given the low star formation rate of the galaxy.

For sources outside of the D25 ellipse of the galaxy, the $\log N (>S)$ versus $\log S$ distribution is roughly consistent with that expected for AGN \citep{bra05} (Figure~\ref{fig:logns}) at fluxes above $2 \times 10^{-14}$~\ergcms. Below this flux, the distribution flattens significantly with decreasing flux. This may be caused by the decrease of sensitivity due to vignetting at large off-axis angles. Some of these sources have bright optical counterparts with an X-ray to optical flux ratio consistent with that of AGN. Therefore, we conclude that most sources outside of the D25 ellipse of  NGC 247 are likely background objects.

\subsection{NGC 247 ULX}

The observed X-ray spectrum of the ULX is completely dominated by a cool, thermal emission component with a temperature around 0.1~keV; a weak power-law tail with a photon index of about 2.5 is also evident but contributes only a small fraction of the flux. This spectral shape is in contrast with that of many other ULXs whose high energy component contributes the majority of the flux (more than 60\%) and can be explained by an optically-thick corona \citep{gla09}. In fact, Gladstone et al. (2009) fits can predict an intrinsically soft-dominated spectrum, but this is a model-dependent result, while the observed spectral shapes of their sources are always dominated by the emission from a hard tail. It is unlikely that the high energy component in NGC 247 ULX arises from an optically thick corona; otherwise, such a corona must have a tiny covering factor since most disk photons are unscattered.

The spectrum, however, is consistent with that of black hole binaries in the thermal state \citep{mcc06,rem06}, and may hint at a connection in nature. An important observational characteristic of the thermal state is that the disk luminosity varies roughly with the 4th power of the disk inner temperature, equivalent to a constant disk inner radius. This source has shown a consistent disk inner radius derived from two XMM-Newton observations, though the constraint from the 2001 observation is loose. For a positive identification of the thermal state, one also needs timing information. This ULX displays red noise with a $\nu^{-1}$ form above the white noise, which is consistent with that seen in the thermal state. However, the integrated rms power is significantly higher than that seen in the thermal state. For Galactic black holes, the total rms power in the 0.1-10~Hz band is usually less than 0.075 in the thermal state, and only in the hard state is the rms power found to be greater than 0.15 \citep{rem06}. However, the X-ray spectrum of the ULX is much too soft to be classified as in the hard state.

If the ULX is in the thermal state, then the source would be an interesting candidate to be an intermediate mass black hole. The accretion disk in the thermal state is believed to extend all the way to the innermost stable circular orbit (ISCO) around the central black hole. The ISCO radius depends only on the black hole mass and spin, as $R_{\rm ISCO} = 6GM/c^2$ for a non-spinning black hole or 6 times smaller for an extremely spinning object, where $G$ is the gravitational constant, $M$ is the black hole mass, and $c$ is the light speed. Assuming a hardening correction of 1.7 and that the maximum temperature occurs at 2.4 times the ISCO radius \citep{mak00}, the ISCO radius can be expressed as $R_{\rm ISCO} \approx 1.2 R_{\rm in}$, where $R_{\rm in}$ is the apparent inner radius derived directly from the normalization of the MCD component. Adopting the parameters from the best-fitted model, MCD plus power-law with absorption edge, and assuming a face-on disk, we have $R_{\rm in} = (0.5-2.2) \times 10^4$~km in the 90\% confidence region, corresponding to the ISCO radius of a $600-3000 M_\sun$ Schwarzschild black hole, or a more massive black hole if spinning or the disk is tilted.  The observed luminosity is about a few percent of the Eddington limit of such a black hole, which is consistent with Eddington ratios typically found in the thermal state. Similar results can be obtained with the Gaussian absorption line.

Due to the lack of a characteristic feature in the power spectrum, we are unable to constrain the black hole mass via timing. The rms-mass relation derived for AGN \citep{zho10} may have issues in extrapolation to lower masses. The faintness in the 2-10 keV band of the source prevents us from implementing the means proposed by \citet{gon11} to estimate the black hole mass.

A few other ULXs show a dominant soft, thermal emission component, and are classified as supersoft ULXs: M101 ULX-1 \citep{pen01,kon05,muk05}, Antennae X-13 \citep{fab03}, M81 ULS1 \citep{swa02,liu08}, and NGC 4631 X1 \citep{car07,sor09}. These sources also show extreme variability: the observed fluxes vary by up to a factor of a few tens, and the intrinsic luminosity changes may be as large as factors of $10^2-10^3$ \citep{kon04,muk05,liu08}. Short-term variability has been detected in almost all of these sources at timescales down to a few ks \citep{swa02,muk05,car07,liu08}. M101 ULX-1 also showed a power spectrum with a $\nu^{-1}$ form at frequencies below $10^{-3}$~Hz \citep{muk05} during its outburst in 2000 March with variability more pronounced at high energies (0.8-2 keV) than in the lower band (0.2-0.8 keV) \citep{muk03}. For a direct comparison, we calculated the rms power of NGC 247 ULX in the same energy bands and found that the fractional rms is 0.80 in the 0.8-2 keV band and 0.44 in the 0.2-0.8 keV band. Similar to NGC 247 ULX, NGC 4631 X1 also exhibits an absorption edge at 1.01~keV with an optical depth of 2.0 in its high state \citep{sor09}. These similarities suggest that NGC 247 ULX may be a new member of the supersoft, ultraluminous family.

The other soft ULXs seem to keep a constant temperature during huge changes in luminosity, indicative of a dramatic change of the emitting surface area of the same order. \citet{fab03} has pointed out that this would lead to difficulty in interpreting the emission as due to accretion power release within a few Schwarzschild radii of the black hole. For NGC 247 ULX, the four ROSAT and two XMM-Newton observations have detected a change by a factor of about 3 of the observed flux in 0.1-2 keV, which is significantly smaller than others. Therefore, the thermal state interpretation has not been ruled out for NGC 247 ULX; new, multiple observations are the key to revealing its nature.

An alternative explanation of the soft, thermal emission is that it arises from a photosphere that has a radius of $10^9$~cm, inferred from the blackbody component. The shortest timescale of variation above white noise is about $10^2$~s (Figure~\ref{fig:pow}), which places a constraint that the emitting area must be smaller than $10^{12}$~cm, compatible with the size obtained from the blackbody parameters. This photosphere is unlikely the surface of a white dwarf, as the observed luminosity and temperature are too high to be explained by steady nuclear burning on a white dwarf surface \citep{sta04}. Plus, presence of the power-law component with a 0.3-10 keV luminosity of $1.2 \times 10^{38}$~\ergs\ is unexpected in such a case. It has been suggested that supercritical accretion could drive massive outflows above the disk and form an optically-thick photosphere, with a temperature expected to be about 0.1 keV, which could shield the inner disk emission when viewed at a high inclination angle \citep{ohs05,pou07}. In this picture, the supersoft ULXs and normal ULXs are the same objects with different viewing angles. This interpretation may be in accord with the absorption feature observed at $\sim$1~keV, which is likely due to L shell transitions of highly ionized iron. Besides this ULX and NGC 4631 X1, less similar absorption features near this energy have been detected in the ULX NGC 1365 X1 \citep{sor07} and in the quasar PDS 456 \citep{ree03}, and are explained as a spectral signature of a massive outflow. The absence of absorption from low-Z elements like oxygen could be a consequence of high ionization.

We conclude that the X-ray properties of the NGC 247 ULX are unlike any popular emission state known in black hole binaries. The source exhibits a spectrum like the thermal state, but shows variability much stronger than seen in the thermal state. A possible strong absorption feature is detected. This source and other supersoft ULXs may exhibit a new accretion state that is different from those seen in Galactic black hole binaries and the majority of ULXs.

A variety of potential thermal emission components have been suggested to be present in the energy spectrum of ULXs, including the soft excess (0.1-0.4~keV) found in the majority of ULXs \citep{fen05,sto06}, the high-temperature ($\sim$keV) ULXs \citep{mak00}, and supersoft ULXs. The properties of these thermal components are summarized and compared in Table~\ref{tab:comparsion}, with NGC 247 ULX and M82 X41.4+60 in the thermal state highlighted. Among them, a disk interpretation is best established for M82 X41.4+60 because its thermal state is consistent with those of stellar-mass X-ray binaries in terms of timing and spectral properties and spectral evolution \citep{fen10}. For the others, definitive evidence is lacking for positive determination of their nature, except that the disk explanation is ruled out for other supersoft ULXs other than NGC 247 ULX. Based on the timing information, it appears unlikely that the soft excess in 'normal' ULXs and the soft (dominant) spectral component in supersoft ULXs have the same physical origin.

\subsection{PHL 6625}

PHL~6625 was identified as a radio-quiet background QSO at $z = 0.38$ based on a Mg~{\sc ii} emission line \citep{mar85}. A single power-law model with interstellar absorption provides an adequate fit to its X-ray spectrum. The parameters listed in Table~\ref{tab:top} are obtained from fitting in the 0.3-10~keV energy range, and the hydrogen column density is very close to the Galactic value at $N_{\rm H} = 2 \times 10^{20}$~cm$^{-2}$, indicative of negligible absorption within NGC 247. However, if we include data down to 0.2~keV, two more energy bins, then the best-fit absorption is $N_{\rm H} = 5.3_{-1.0}^{+1.2} \times 10^{20}$~cm$^{-2}$. This is caused by the low flux between 0.2-0.3~keV, and is consistent with the results obtained from ROSAT observations, in which a total absorption column density $N_{\rm H} = 5.4_{-2.1}^{+2.4} \times 10^{20}$~cm$^{-2}$ was derived from fitting in the 0.1-2~keV band, and the excess absorption mainly takes effect at energies below 0.3~keV \citep{elv97}. Those authors claimed a positive detection of absorbing materials in NGC 247 assuming a Galactic $N_{\rm H} = 1.4 \times 10^{20}$~cm$^{-2}$ based on 21~cm observations of neutral hydrogen. Here, we confirm this conclusion with tighter constraints, even assuming a higher Galactic absorption. Therefore, with future high-resolution spectroscopic observations, this QSO could be used to study the interstellar medium within NGC 247.


For the 2009 data, the power-law photon index at the rest frame is $\Gamma = 2.10_{-0.09}^{+0.13}$. We extracted spectra for this object from the 2001 XMM-Newton observation and found a harder spectrum with $\Gamma = 1.64 \pm 0.17$ at a flux level of $3.7 \pm 0.6 \times 10^{-13}$~\ergcms\ in 0.3-10~keV similar to that in 2009. The fluxes observed with XMM-Newton are about ten times lower than the ROSAT flux of $(5.9 \pm 0.7) \times 10^{-12}$~\ergcms\ in the 0.2-2.4 keV band \citep{elv97}. No short-term X-ray variability has been detected for this object.

\subsection{Other individual sources}

{\bf Source 28}. --- The source is hard, and its spectrum can be described by a power-law with $\Gamma = 1.27$, but the residuals vary systematically with energy. Adding an MCD component improves the residuals and results in $\chi^2 / {\rm dof} = 31.7/54$. The disk has a temperature at the inner radius of $0.35_{-0.10}^{+0.14}$~keV with a bolometric luminosity of $2 \times 10^{37}$~\ergs\ assuming a face-on geometry, and the power-law component has a photon index $\Gamma = 0.6_{-0.3}^{+0.2}$. The disk component has a fractional luminosity of 11\% in the 0.3-10 keV band. If this source is a black hole binary, it is possibly in the hard state.


{\bf Source 34}. --- The source sits at the north end of the NGC 247 disk, and was bright during the 2001 XMM-Newton observation with $\Gamma = 2.3_{-0.9}^{+1.5}$, $N_{\rm H} = 1.9_{-0.8}^{+2.4} \times 10^{21}$~cm$^{-2}$, and $f_{\rm X} = 2.0 \times 10^{-13}$~\ergcms\ in the 0.3-10 keV band, or $L_{\rm X} = 4.6 \times 10^{38}$~\ergs\ corrected for absorption. However, it became faint in the 2009 XMM-Newton observation, with a flux of $3.4 \times 10^{-14}$~\ergcms\ in the 0.3-10 keV band, six times dimmer than in 2001. It is possibly source 3 in \citet{fab92} from Einstein observations and source 1 in \citet{rea97} from ROSAT observations. The ROSAT observations revealed a luminosity of $1.7 \times 10^{38}$~\ergs\ corrected for Galactic absorption only (at 3.4~Mpc) in the 0.1-2 keV band, or $3 \times 10^{38}$~\ergs\ corrected for all absorption. The luminosity in the same band from the 2001 XMM-Newton observation is about $6 \times 10^{38}$~\ergs.

There is no bright optical counterpart found at the X-ray position. However, the source is associated with a possible star forming region, seen in the ultraviolet image from the 2001 XMM-Newton observation, with diffuse emission and star clusters nearby. Therefore, the source is likely an X-ray binary in NGC 247, and is perhaps an accreting black hole.

{\bf Source 36}. --- This source could be source 3 in \citet{rea97}. The best-fitted column density is consistent with the Galactic value. Its flux estimated from the 2001 XMM-Newton observation is consistent with that in 2009. In the X-ray image, the source seems to show extended features and may consists of multiple objects, which can only be resolved with future Chandra observations.

{\bf Source 61}. --- Its X-ray spectrum cannot be adequately fitted by a single power-law model, with $\chi^2 / {\rm dof} = 80.9/60$; adding a Gaussian absorption line significantly improves the fit with $\chi^2 / {\rm dof} = 58.1/57$. The absorption line has a central energy of $0.9 \pm 0.1$~keV and a width of $\sigma = 0.22_{-0.16}^{+0.28}$~keV with an optical depth of $0.29_{-0.19}^{+0.58}$. The equivalent width is calculated to be around 0.30~keV. Its observed flux in 2001 was about twice of that in 2009. There is a bright optical source with $B = 18.83$ and $R = 18.58$ within 2\arcsec\ of the X-ray position, and the X-ray to optical flux ratio is estimated to be $-0.3$. As discussed above, this source is likely a background AGN. Due to the low absorption column density, the absorption line is likely produced intrinsic to the source. Similar absorption has been seen in AGN \citep{blu02}.

{\bf Source 13}. --- This is the hardest source detected with XMM-Newton. It is obvious in the hard band (2-10~keV) but undetected in the soft band (0.3-2~keV), consistent with $N_{\rm H} > 5 \times 10^{22}$~cm$^{-2}$ for a power-law spectrum with $\Gamma = 2$. The source lies about 7\arcmin\ west of the nucleus of NGC 247, and is likely a highly absorbed background AGN.

{\bf Sources 41, 50, 59, and 72}. --- These four sources are the softest besides the ULX and are only detected in the soft band (0.3-2 keV) with a hardness ratio around $-0.9$, which is consistent with a spectrum with a blackbody temperature $< 0.4$~keV or a power-law photon index $\Gamma > 2.5$, assuming Galactic absorption.  Sources 41 and 72 are outside of the D25 ellipse of NGC 247, but sources 50 and 59 are inside. Their X-ray luminosities are about $10^{36} - 10^{37}$~\ergs. These sources do not appear to have bright optical counterparts. They could be canonical supersoft sources that are powered by nuclear burning on the surface of white dwarfs, or high redshift AGN.


{\it Facility:} \facility{XMM-Newton}

\LongTables

\begin{deluxetable}{llccccccccc}
\tablecolumns{11}
\setlength{\tabcolsep}{3pt}
\tabletypesize{\footnotesize}
\tablewidth{\textwidth}
\tablecaption{The XMM-Newton source list of NGC 247.
\label{tab:src}}
\tablehead{
\colhead{No.} & \colhead{IAU name} & \colhead{R.A.} & \colhead{decl.} & \colhead{err} & \colhead{flux1} & \colhead{flux2} & \colhead{hardness} & \colhead{like.} & \colhead{D25} & \colhead{note} \\
\colhead{} & \colhead{(XMMU)} & \colhead{(J2000.0)} & \colhead{(J2000.0)} & \colhead{(\arcsec)} & \colhead{(0.3-2 keV)} & \colhead{(2-10 keV)} & \colhead{} & \colhead{} & \colhead{} & \colhead{} \\
\colhead{(1)} & \colhead{(2)} & \colhead{(3)} & \colhead{(4)} & \colhead{(5)} & \colhead{(6)} & \colhead{(7)} & \colhead{(8)} & \colhead{(9)} & \colhead{(10)} & \colhead{(11)}
}
\startdata
1 & J004609.2$-$204504 & 00 46 09.29 & $-$20 45 04.9 &  2.21 & $0.050\pm0.014$ & $0.09\pm0.07$ & $-0.5\pm0.3$ &    13.7 & no &  \\
2 & J004617.0$-$204151 & 00 46 17.00 & $-$20 41 51.8 &  1.44 & $0.021\pm0.013$ & $0.27\pm0.08$ & $0.6\pm0.3$ &    15.9 & no &  \\
3 & J004623.0$-$204740 & 00 46 23.02 & $-$20 47 40.3 &  1.02 & $0.066\pm0.012$ & $0.21\pm0.07$ & $-0.26\pm0.18$ &    73.5 & no &  \\
4 & J004628.7$-$205005 & 00 46 28.73 & $-$20 50 05.2 &  0.78 & $0.22\pm0.02$ & $0.26\pm0.06$ & $-0.54\pm0.09$ &   195.4 & no & O \\
5 & J004631.1$-$203959 & 00 46 31.14 & $-$20 39 59.1 &  1.89 & $0.052\pm0.015$ & $0.25\pm0.08$ & $0.1\pm0.2$ &    17.2 & no &  \\
6 & J004632.1$-$205044 & 00 46 32.11 & $-$20 50 44.2 &  1.85 & $0.032\pm0.009$ & $0.13\pm0.06$ & $-0.2\pm0.3$ &    18.6 & no &  \\
7 & J004633.1$-$205457 & 00 46 33.16 & $-$20 54 57.3 &  1.40 & $0.065\pm0.014$ & $0.06\pm0.04$ & $-0.7\pm0.2$ &    20.1 & no &  \\
8 & J004634.1$-$205411 & 00 46 34.15 & $-$20 54 11.2 &  0.93 & $0.042\pm0.009$ & $0.08\pm0.05$ & $-0.5\pm0.2$ &    51.1 & no &  \\
9 & J004635.1$-$203529 & 00 46 35.17 & $-$20 35 29.2 &  1.25 & $0.17\pm0.04$ & $0.37\pm0.16$ & $-0.2\pm0.2$ &    42.3 & no &  \\
10 & J004635.2$-$203648 & 00 46 35.23 & $-$20 36 48.2 &  1.43 & $0.086\pm0.018$ & $0.35\pm0.12$ & $-0.10\pm0.19$ &    40.6 & no &  \\
11 & J004636.2$-$210018 & 00 46 36.29 & $-$21 00 18.7 &  1.00 & $0.22\pm0.03$ & $0.11\pm0.07$ & $-0.75\pm0.14$ &    99.8 & no & O \\
12 & J004638.0$-$203609 & 00 46 38.04 & $-$20 36 09.6 &  1.62 & $0.079\pm0.018$ & $0.08\pm0.09$ & $-0.7\pm0.3$ &    32.5 & no &  \\
13 & J004639.6$-$204533 & 00 46 39.67 & $-$20 45 33.6 &  1.05 & $<0.0047$ & $0.26\pm0.06$ & $>0.79$ &    40.4 & no &  \\
14 & J004642.6$-$204507 & 00 46 42.64 & $-$20 45 07.5 &  1.49 & $0.032\pm0.008$ & $0.05\pm0.04$ & $-0.6\pm0.3$ &    19.4 & no &  \\
15 & J004645.8$-$210030 & 00 46 45.85 & $-$21 00 30.6 &  1.78 & $0.084\pm0.019$ & $0.19\pm0.08$ & $-0.2\pm0.2$ &    21.7 & no &  \\
16 & J004646.3$-$204742 & 00 46 46.39 & $-$20 47 42.7 &  0.51 & $0.112\pm0.012$ & $0.34\pm0.06$ & $-0.30\pm0.09$ &   327.2 & no & O \\
17 & J004646.9$-$203658 & 00 46 46.92 & $-$20 36 58.0 &  1.56 & $0.063\pm0.015$ & $0.13\pm0.07$ & $-0.4\pm0.3$ &    19.8 & no &  \\
18 & J004647.2$-$203628 & 00 46 47.23 & $-$20 36 28.1 &  1.78 & $0.070\pm0.016$ & $0.14\pm0.08$ & $-0.4\pm0.3$ &    22.1 & no &  \\
19 & J004647.7$-$203837 & 00 46 47.72 & $-$20 38 37.3 &  0.65 & $0.23\pm0.02$ & $0.37\pm0.09$ & $-0.52\pm0.09$ &   333.1 & no & O \\
20 & J004648.8$-$205247 & 00 46 48.85 & $-$20 52 47.0 &  1.20 & $0.053\pm0.009$ & $0.07\pm0.04$ & $-0.60\pm0.19$ &    60.6 & no &  \\
21 & J004650.4$-$203854 & 00 46 50.47 & $-$20 38 54.0 &  1.76 & $0.042\pm0.012$ & $0.05\pm0.06$ & $-0.6\pm0.4$ &    17.4 & no &  \\
22 & J004651.8$-$204328 & 00 46 51.82 & $-$20 43 28.5 &  0.14 & $2.26\pm0.05$ & $2.62\pm0.14$ & $-0.655\pm0.016$ & 14105.3 & no & QSO \\
23 & J004654.6$-$204129 & 00 46 54.63 & $-$20 41 29.4 &  0.78 & $0.086\pm0.012$ & $0.12\pm0.05$ & $-0.59\pm0.15$ &    92.5 & yes & O \\
24 & J004654.8$-$204100 & 00 46 54.87 & $-$20 41 00.9 &  0.56 & $0.196\pm0.018$ & $0.42\pm0.08$ & $-0.43\pm0.08$ &   381.1 & yes &  \\
25 & J004655.7$-$203407 & 00 46 55.74 & $-$20 34 07.6 &  1.55 & $0.084\pm0.018$ & $0.12\pm0.12$ & $-0.5\pm0.3$ &    19.5 & no &  \\
26 & J004655.9$-$205248 & 00 46 55.92 & $-$20 52 48.4 &  1.48 & $0.026\pm0.008$ & $0.11\pm0.04$ & $0.0\pm0.2$ &    18.1 & no &  \\
27 & J004656.4$-$205611 & 00 46 56.43 & $-$20 56 11.9 &  1.34 & $0.053\pm0.010$ & $0.19\pm0.06$ & $-0.20\pm0.18$ &    35.7 & no &  \\
28 & J004659.3$-$204538 & 00 46 59.38 & $-$20 45 38.8 &  0.26 & $0.336\pm0.019$ & $1.06\pm0.09$ & $-0.28\pm0.05$ &  1796.1 & yes &  \\
29 & J004659.7$-$204227 & 00 46 59.70 & $-$20 42 27.3 &  1.34 & $0.063\pm0.010$ & $0.03\pm0.04$ & $-0.84\pm0.17$ &    38.7 & yes &  \\
30 & J004700.5$-$203359 & 00 47 00.55 & $-$20 33 59.7 &  3.09 & $0.05\pm0.02$ & $0.35\pm0.11$ & $0.3\pm0.2$ &    10.9 & no &  \\
31 & J004701.9$-$205450 & 00 47 01.98 & $-$20 54 50.8 &  1.62 & $0.026\pm0.007$ & $0.08\pm0.04$ & $-0.3\pm0.3$ &    23.2 & no &  \\
32 & J004702.7$-$205302 & 00 47 02.79 & $-$20 53 02.9 &  1.97 & $0.032\pm0.010$ & $0.03\pm0.03$ & $-0.7\pm0.3$ &    14.7 & yes &  \\
33 & J004703.2$-$204545 & 00 47 03.23 & $-$20 45 45.9 &  1.12 & $0.042\pm0.009$ & $0.17\pm0.04$ & $-0.01\pm0.18$ &    56.2 & yes &  \\
34 & J004703.2$-$203657 & 00 47 03.28 & $-$20 36 57.9 &  1.64 & $0.034\pm0.011$ & $0.31\pm0.09$ & $0.3\pm0.2$ &    20.1 & yes &  \\
35 & J004703.4$-$204321 & 00 47 03.44 & $-$20 43 21.5 &  1.02 & $0.034\pm0.009$ & $0.21\pm0.05$ & $0.20\pm0.18$ &    42.7 & yes &  \\
36 & J004703.6$-$204015 & 00 47 03.68 & $-$20 40 15.6 &  0.42 & $0.43\pm0.03$ & $0.60\pm0.09$ & $-0.59\pm0.05$ &  1184.2 & yes &  \\
37 & J004703.9$-$204743 & 00 47 03.90 & $-$20 47 43.8 &  0.16 & $1.84\pm0.04$ & $0.13\pm0.04$ & $-0.975\pm0.008$ &  9827.5 & yes & ULX \\
38 & J004704.5$-$205812 & 00 47 04.57 & $-$20 58 12.0 &  1.94 & $0.061\pm0.019$ & $0.23\pm0.09$ & $-0.0\pm0.2$ &    17.8 & no &  \\
39 & J004705.9$-$205239 & 00 47 05.92 & $-$20 52 39.4 &  0.75 & $0.071\pm0.010$ & $0.11\pm0.04$ & $-0.57\pm0.14$ &   152.1 & yes & O \\
40 & J004705.9$-$204152 & 00 47 05.94 & $-$20 41 52.4 &  1.07 & $0.088\pm0.013$ & $0.13\pm0.06$ & $-0.57\pm0.15$ &    69.9 & yes &  \\
41 & J004707.3$-$205749 & 00 47 07.38 & $-$20 57 49.3 &  1.30 & $0.073\pm0.013$ & $<0.04$ & $<-0.78$ &    51.4 & no &  \\
42 & J004708.0$-$204346 & 00 47 08.01 & $-$20 43 46.8 &  1.57 & $0.020\pm0.008$ & $0.13\pm0.04$ & $0.2\pm0.3$ &    10.4 & yes &  \\
43 & J004709.0$-$205214 & 00 47 09.03 & $-$20 52 14.7 &  0.73 & $0.077\pm0.011$ & $0.30\pm0.06$ & $-0.18\pm0.12$ &   172.1 & yes &  \\
44 & J004709.1$-$204622 & 00 47 09.16 & $-$20 46 22.8 &  1.10 & $0.060\pm0.010$ & $0.19\pm0.04$ & $-0.22\pm0.13$ &    56.1 & yes &  \\
45 & J004710.0$-$204708 & 00 47 10.06 & $-$20 47 08.4 &  0.34 & $0.237\pm0.016$ & $0.81\pm0.08$ & $-0.25\pm0.05$ &  1117.2 & yes &  \\
46 & J004710.3$-$204343 & 00 47 10.32 & $-$20 43 43.9 &  1.29 & $0.029\pm0.008$ & $0.07\pm0.04$ & $-0.4\pm0.3$ &    24.8 & yes &  \\
47 & J004712.4$-$204219 & 00 47 12.48 & $-$20 42 19.1 &  1.45 & $0.027\pm0.009$ & $0.10\pm0.04$ & $-0.2\pm0.4$ &    15.6 & yes &  \\
48 & J004712.7$-$203837 & 00 47 12.74 & $-$20 38 37.0 &  1.88 & $0.042\pm0.011$ & $0.15\pm0.07$ & $-0.2\pm0.3$ &    14.0 & yes & O, I \\
49 & J004713.7$-$205113 & 00 47 13.70 & $-$20 51 13.4 &  1.15 & $0.033\pm0.008$ & $0.09\pm0.04$ & $-0.4\pm0.2$ &    45.1 & yes &  \\
50 & J004715.1$-$204717 & 00 47 15.17 & $-$20 47 17.3 &  0.96 & $0.059\pm0.009$ & $0.01\pm0.02$ & $-0.92\pm0.11$ &    66.3 & yes &  \\
51 & J004715.9$-$205502 & 00 47 15.94 & $-$20 55 02.5 &  0.75 & $0.161\pm0.017$ & $0.21\pm0.06$ & $-0.61\pm0.10$ &   235.8 & yes &  \\
52 & J004716.5$-$204814 & 00 47 16.59 & $-$20 48 14.4 &  0.85 & $0.047\pm0.008$ & $0.10\pm0.04$ & $-0.47\pm0.17$ &    84.1 & yes &  \\
53 & J004716.6$-$203346 & 00 47 16.63 & $-$20 33 46.4 &  1.09 & $0.056\pm0.018$ & $0.39\pm0.14$ & $0.2\pm0.2$ &    11.1 & no &  \\
54 & J004717.9$-$205508 & 00 47 17.93 & $-$20 55 08.6 &  0.99 & $0.111\pm0.014$ & $0.23\pm0.06$ & $-0.45\pm0.12$ &    97.5 & yes &  \\
55 & J004718.5$-$204739 & 00 47 18.56 & $-$20 47 39.8 &  0.90 & $0.072\pm0.011$ & $0.17\pm0.04$ & $-0.38\pm0.15$ &    64.2 & yes &  \\
56 & J004719.7$-$205601 & 00 47 19.77 & $-$20 56 01.3 &  1.90 & $0.032\pm0.009$ & $0.07\pm0.05$ & $-0.4\pm0.3$ &    11.8 & no &  \\
57 & J004720.3$-$205248 & 00 47 20.32 & $-$20 52 48.7 &  1.00 & $0.083\pm0.012$ & $0.06\pm0.04$ & $-0.78\pm0.16$ &   110.1 & yes & O($2$), I \\
58 & J004721.0$-$204932 & 00 47 21.05 & $-$20 49 32.3 &  1.32 & $0.038\pm0.009$ & $0.13\pm0.04$ & $-0.1\pm0.2$ &    25.0 & yes &  \\
59 & J004721.4$-$205104 & 00 47 21.48 & $-$20 51 04.1 &  1.80 & $0.021\pm0.007$ & $<0.012$ & $<-0.80$ &    14.8 & yes &  \\
60 & J004725.2$-$205442 & 00 47 25.23 & $-$20 54 42.4 &  1.28 & $0.15\pm0.03$ & $0.13\pm0.07$ & $-0.64\pm0.17$ &    67.5 & no &  \\
61 & J004727.3$-$204721 & 00 47 27.36 & $-$20 47 21.6 &  0.28 & $0.58\pm0.03$ & $0.55\pm0.08$ & $-0.71\pm0.04$ &  2204.7 & no & O \\
62 & J004728.7$-$204301 & 00 47 28.76 & $-$20 43 01.7 &  0.72 & $0.127\pm0.016$ & $0.21\pm0.07$ & $-0.54\pm0.12$ &   162.3 & no & O \\
63 & J004730.6$-$205434 & 00 47 30.64 & $-$20 54 34.0 &  0.68 & $0.23\pm0.03$ & $0.56\pm0.12$ & $-0.23\pm0.12$ &   223.0 & no &  \\
64 & J004733.5$-$204557 & 00 47 33.57 & $-$20 45 57.5 &  1.16 & $0.035\pm0.009$ & $0.18\pm0.06$ & $-0.0\pm0.2$ &    27.4 & no &  \\
65 & J004737.4$-$203648 & 00 47 37.49 & $-$20 36 48.3 &  1.50 & $0.060\pm0.017$ & $0.39\pm0.13$ & $0.1\pm0.2$ &    25.3 & no & O \\
66 & J004740.7$-$204442 & 00 47 40.77 & $-$20 44 42.5 &  1.87 & $0.040\pm0.011$ & $0.15\pm0.08$ & $-0.2\pm0.3$ &    15.9 & no &  \\
67 & J004741.8$-$205231 & 00 47 41.84 & $-$20 52 31.5 &  0.52 & $0.38\pm0.03$ & $0.79\pm0.11$ & $-0.30\pm0.07$ &   485.8 & no & O($2$) \\
68 & J004743.4$-$204941 & 00 47 43.49 & $-$20 49 41.4 &  0.68 & $0.178\pm0.019$ & $0.29\pm0.08$ & $-0.53\pm0.11$ &   230.5 & no &  \\
69 & J004744.5$-$205321 & 00 47 44.53 & $-$20 53 21.2 &  1.03 & $0.12\pm0.02$ & $0.24\pm0.08$ & $-0.31\pm0.17$ &    65.2 & no & O \\
70 & J004752.1$-$204053 & 00 47 52.19 & $-$20 40 53.9 &  1.78 & $0.066\pm0.017$ & $0.12\pm0.12$ & $-0.5\pm0.4$ &    22.5 & no & O \\
71 & J004753.5$-$205435 & 00 47 53.55 & $-$20 54 35.3 &  1.23 & $0.11\pm0.02$ & $0.09\pm0.06$ & $-0.5\pm0.3$ &    39.8 & no &  \\
72 & J004759.1$-$205201 & 00 47 59.19 & $-$20 52 01.2 &  2.02 & $0.068\pm0.017$ & $0.03\pm0.06$ & $-0.9\pm0.3$ &    11.7 & no &  \\
73 & J004808.3$-$204553 & 00 48 08.39 & $-$20 45 53.3 &  3.61 & $0.053\pm0.019$ & $0.53\pm0.15$ & $0.3\pm0.2$ &    13.7 & no &  \\
74 & J004810.7$-$204818 & 00 48 10.75 & $-$20 48 18.5 &  9.07 & $0.029\pm0.016$ & $0.57\pm0.16$ & $0.6\pm0.2$ &    11.2 & no &  \\
75 & J004812.3$-$204755 & 00 48 12.30 & $-$20 47 55.6 &  2.56 & $0.031\pm0.016$ & $0.54\pm0.15$ & $0.6\pm0.2$ &    12.4 & no &  \\
\enddata
\tablecomments{
Col.~(2-5): Source names, coordinates, and position errors.
Col.~(6-7): Absorbed fluxes in units of $10^{-13}$~\ergcms.
Col.~(8-10): Hardness ratios, detection likelihood, and whether or not the source is inside the D25 ellipse of NGC 247.
Col.~(11): `O' denotes optical counterparts found in USNO-B1.0 with numbers, if more than one, shown in parentheses; `I' denotes counterparts found in 2MASS. 90\% upper bounds are quoted for non-detections. }
\end{deluxetable}


\begin{deluxetable}{lllllll}
\tablewidth{0pc}
\tablecaption{Spectral parameters of bright sources in NGC 247 fitted with a power-law model.
\label{tab:top}}
\tablehead{
\colhead{No.} & \colhead{$N_{\rm H}$} & \colhead{$\Gamma$} & \colhead{$N_{\rm PL}$} & \colhead{$f_{\rm X}$} & \colhead{$L_{\rm X}$} &  \colhead{$\chi^2$/dof}
}
\startdata
4 & $0.2_{-0}^{+0.8}$ & $1.7_{-0.3}^{+0.4}$ & $1.01_{-0.17}^{+0.43}$ & $0.71_{-0.22}^{+0.15}$ & $1.0_{-0.2}^{+0.3}$ & 8.7/10 \\
16 & $0.2_{-0}^{+0.7}$ & $1.2_{-0.3}^{+0.4}$ & $0.45_{-0.08}^{+0.15}$ & $0.52_{-0.16}^{+0.13}$ & $0.7_{-0.2}^{+0.2}$ & 20.8/17 \\
19 & $0.2_{-0}^{+0.3}$ & $2.0_{-0.3}^{+0.3}$ & $0.90_{-0.12}^{+0.14}$ & $0.50_{-0.14}^{+0.09}$ & $0.70_{-0.14}^{+0.17}$ & 18.3/15 \\
22\tablenotemark{*} & $0.27_{-0.07}^{+0.17}$ & $2.10_{-0.09}^{+0.13}$ & $17.4_{-1.3}^{+2.1}$ & $4.3_{-0.3}^{+0.3}$ & $2.26_{-0.12}^{+0.12}\times10^6$ & 99.2/99 \\
24 & $1.9_{-0.8}^{+1.1}$ & $2.3_{-0.2}^{+0.6}$ & $1.4_{-0.5}^{+1.0}$ & $0.41_{-0.11}^{+0.16}$ & $1.0_{-0.2}^{+0.8}$ & 17.9/14 \\
28 & $0.20_{-0}^{+0.08}$ & $1.27_{-0.10}^{+0.10}$ & $1.42_{-0.10}^{+0.10}$ & $1.55_{-0.14}^{+0.13}$ & $2.16_{-0.19}^{+0.20}$ & 57.2/56 \\
36 & $0.2_{-0}^{+0.2}$ & $1.69_{-0.11}^{+0.14}$ & $2.07_{-0.14}^{+0.22}$ & $1.43_{-0.23}^{+0.15}$ & $2.0_{-0.2}^{+0.3}$ & 59.9/52 \\
39 & $0.2_{-0}^{+0.8}$ & $1.5_{-0.4}^{+0.6}$ & $0.34_{-0.06}^{+0.19}$ & $0.30_{-0.14}^{+0.11}$ & $0.42_{-0.14}^{+0.16}$ & 7.3/8 \\
43 & $3.9_{-1.6}^{+2.6}$ & $1.9_{-0.2}^{+0.3}$ & $0.8_{-0.4}^{+0.9}$ & $0.34_{-0.10}^{+0.12}$ & $0.71_{-0.15}^{+0.42}$ & 6.0/11 \\
45 & $1.5_{-0.5}^{+0.5}$ & $1.66_{-0.16}^{+0.16}$ & $1.7_{-0.3}^{+0.4}$ & $1.02_{-0.14}^{+0.15}$ & $1.72_{-0.17}^{+0.17}$ & 46.8/46 \\
51 & $1.2_{-0.8}^{+0.9}$ & $1.9_{-0.5}^{+0.3}$ & $0.8_{-0.3}^{+0.5}$ & $0.37_{-0.11}^{+0.15}$ & $0.67_{-0.13}^{+0.23}$ & 7.6/8 \\
61 & $0.20_{-0}^{+0.06}$ & $2.13_{-0.13}^{+0.13}$ & $2.16_{-0.13}^{+0.13}$ & $1.13_{-0.08}^{+0.09}$ & $1.58_{-0.14}^{+0.20}$ & 80.9/60 \\
62 & $0.2_{-0}^{+0.5}$ & $1.8_{-0.4}^{+0.5}$ & $0.59_{-0.10}^{+0.15}$ & $0.38_{-0.15}^{+0.11}$ & $0.51_{-0.09}^{+0.20}$ & 12.2/12 \\
63 & $2.2_{-1.3}^{+1.5}$ & $1.8_{-0.2}^{+0.4}$ & $1.4_{-0.5}^{+0.9}$ & $0.7_{-0.2}^{+0.3}$ & $1.3_{-0.2}^{+0.3}$ & 7.2/13 \\
67 & $0.6_{-0.4}^{+1.0}$ & $1.7_{-0.4}^{+0.4}$ & $1.8_{-0.5}^{+0.9}$ & $1.2_{-0.3}^{+0.3}$ & $1.8_{-0.3}^{+0.4}$ & 14.1/13 \\
68 & $0.2_{-0}^{+0.8}$ & $1.8_{-0.3}^{+0.6}$ & $0.78_{-0.11}^{+0.36}$ & $0.49_{-0.18}^{+0.15}$ & $0.70_{-0.17}^{+0.21}$ & 19.3/14 \\
\enddata

\tablecomments{$N_{\rm H}$ is the absorption column density in ${10^{21}}$~cm$^{-2}$. $\Gamma$ is the power-law photon index. $N_{\rm PL}$ is the power-law normalization at 1~keV in units of $10^{-5}$~photons~keV$^{-1}$~cm$^{-2}$~s$^{-1}$. $f_{\rm X}$ is the observed flux in 0.3-10 keV in units of $10^{-13}$~\ergcms. $L_{\rm X}$ is the unabsorbed luminosity in 0.3-10 keV in units of $10^{38}$~\ergs. Errors are quoted at 90\% confidence level. }
\tablenotetext{*}{A redshifted power-law model with $z = 0.38$ is used, and a luminosity distance of 2.0~Gpc assuming $H_0 = 70$~km~s$^{-1}$~Mpc$^{-1}$ is adopted.}
\end{deluxetable}

\begin{deluxetable}{llllcllcllllll}
\tablecaption{Spectral parameters of NGC 247 ULX.
\label{tab:spe}}
\setlength{\tabcolsep}{3pt}
\tabletypesize{\footnotesize}
\tablewidth{0pc}
\tablehead{
\colhead{} & \colhead{} & \multicolumn{2}{c}{thermal} & \colhead{} & \multicolumn{2}{c}{power-law} & \colhead{} & \multicolumn{3}{c}{absorption} & \colhead{} & \colhead{} & \colhead{} \\
\cline{3-4} \cline{6-7} \cline{9-11} \noalign{\smallskip}
\colhead{} & \colhead{$N_{\rm H}$} & \colhead{$kT$} & \colhead{$R$} & \colhead{} & \colhead{$\Gamma$} & \colhead{$N_{\rm PL}$} & \colhead{} & \colhead{$E$} & \colhead{$\tau$} & \colhead{$\sigma$} & \colhead{$f_{\rm X}$} & \colhead{$L_{\rm X}$} & \colhead{$\chi^2$/dof}
}
\startdata
$\bf a_1$ & 3.2\tablenotemark{*} & $0.117_{-0.003}^{+0.003}$ & $1.34_{-0.13}^{+0.12}$ && $1.6_{-0.7}^{+0.6}$ & $0.5_{-0.3}^{+0.5}$ && \nodata & \nodata & \nodata & $1.75_{-0.15}^{+0.18}$ & $2.92_{-0.16}^{+0.16}$ & 88.0/43 \\
$\bf a_2$ & $3.2_{-0.7}^{+0.7}$ & $0.13_{-0.02}^{+0.02}$ & $0.9_{-0.5}^{+1.3}$ && $2.5_{-1.0}^{+1.4}$ & $1.7_{-1.2}^{+3.8}$ && $1.02_{-0.03}^{+0.02}$ & $1.4_{-0.5}^{+0.5}$ & \nodata & $1.71_{-0.13}^{+0.17}$ & $2.7_{-1.3}^{+3.4}$ & 36.8/40 \\
$\bf a_3$ & $3.5_{-0.8}^{+0.9}$ & $0.123_{-0.012}^{+0.022}$ & $1.3_{-0.4}^{+1.2}$ && $2.0_{-1.0}^{+1.1}$ & $0.9_{-0.6}^{+1.3}$ && $1.16_{-0.03}^{+0.05}$ & $0.26_{-0.08}^{+0.08}$ & $0.08_{-0.03}^{+0.04}$ & $1.74_{-0.14}^{+0.18}$ & $3.4_{-1.6}^{+4.0}$ & 37.0/39 \\
\noalign{\smallskip}\hline\noalign{\smallskip}
$\bf b_1$ & 2.7\tablenotemark{*} & $0.103_{-0.002}^{+0.003}$ & $1.35_{-0.11}^{+0.07}$ && $1.6_{-0.7}^{+0.6}$ & $0.6_{-0.3}^{+0.5}$ && \nodata & \nodata & \nodata & $1.76_{-0.15}^{+0.18}$ & $1.74_{-0.09}^{+0.09}$ & 81.5/43 \\
$\bf b_2$ & $2.7_{-0.7}^{+1.0}$ & $0.112_{-0.017}^{+0.016}$ & $1.0_{-0.5}^{+1.3}$ && $2.7_{-1.0}^{+2.2}$ & $2.0_{-1.4}^{+7.5}$ && $1.02_{-0.03}^{+0.03}$ & $1.3_{-0.4}^{+0.5}$ & \nodata & $1.70_{-0.14}^{+0.17}$ & $1.6_{-0.8}^{+2.9}$ & 39.0/40 \\
$\bf b_3$ & $2.9_{-0.7}^{+0.9}$ & $0.109_{-0.013}^{+0.017}$ & $1.2_{-0.5}^{+1.2}$ && $2.1_{-0.9}^{+1.2}$ & $1.0_{-0.6}^{+1.4}$ && $1.15_{-0.03}^{+0.04}$ & $0.23_{-0.08}^{+0.07}$ & $0.08_{-0.03}^{+0.04}$ & $1.74_{-0.14}^{+0.18}$ & $1.9_{-0.9}^{+1.7}$ & 38.9/39
\enddata
\tablecomments{
XSPEC models $\bf a_1$: {\tt wabs $\ast$ (diskbb + powerlaw)};
$\bf a_2$: {\tt wabs $\ast$ edge $\ast$ (diskbb + powerlaw)};
$\bf a_3$: {\tt wabs $\ast$ gabs $\ast$ (diskbb + powerlaw)};
$\bf b_1$: {\tt wabs $\ast$ (bbodyrad + powerlaw)};
$\bf b_2$: {\tt wabs $\ast$ edge $\ast$ (bbodyrad + powerlaw)};
$\bf b_3$: {\tt wabs $\ast$ gabs $\ast$ (bbodyrad + powerlaw)}.
$N_{\rm H}$ is the absorption column density in ${10^{21}}$~cm$^{-2}$. $kT$ and $R$ are temperature in keV and face-on radius in $10^4$~km of the thermal component, respectively; for the {\tt diskbb} model, they correspond to the values at the innermost disk. $\Gamma$ is the power-law photon index. $N_{\rm PL}$ is the power-law normalization at 1~keV in units of $10^{-5}$~photons~keV$^{-1}$~cm$^{-2}$~s$^{-1}$. $E$ and $\tau$ are energy in keV and optical depth of the local absorption component, respectively, and $\sigma$ is the width of the Gaussian absorption in keV. $f_{\rm X}$ is the observed flux in 0.3-10 keV in units of $10^{-13}$~\ergcms. $L_{\rm X}$ is the unabsorbed luminosity in 0.3-10 keV in units of $10^{39}$~\ergs. Errors are quoted at 90\% confidence level.}
\tablenotetext{*}{Fixed at the value obtained from the best-fit model.}
\end{deluxetable}

\begin{deluxetable}{llccc}
\tablewidth{0pc}
\tablecaption{Comparison of the thermal emission component in ULXs.
\label{tab:comparsion}}
\tablehead{
 & \colhead{Temperature} & \colhead{Fraction} & \colhead{Fast Variability} & \colhead{$L\propto T^4$}
}
\startdata
NGC 247 ULX & low ($\sim$0.1 keV)  & major & high  & unknown \\
other supersoft ULXs  & low ($\sim$0.1 keV) & major & high & no \\
M82 X41.4+60 during outburst\tablenotemark{a} & high ($\sim$1 keV) & major & low & yes \\
other hot ULXs & high ($\sim$1 keV) & major & unknown &  maybe\tablenotemark{b} \\
soft excesses in ULXs & low (0.1-0.4 keV) & minor & weak\tablenotemark{c} & rare\tablenotemark{d} \\
\enddata
\tablecomments{Temperature is the typical temperature of the thermal emission component found in the X-ray spectrum. Fraction is `major' if the thermal emission component produces more than 60\% of the total X-ray flux. Fast Variability is `low' if the variability of the thermal emission component is consistent with that of white noise. The last column indicates if the luminosity varies as the fourth power of the temperature of the thermal emission component.}
\tablenotetext{a}{when it is in the thermal dominant state \citep{fen10}.}
\tablenotetext{b}{possibly seen in M82 X37.8+54 \citep{jin10}, Suzaku J1305$-$4931 \citep{iso08}, and NGC 253 X-2 \citep{kaj09}}
\tablenotetext{c}{A few ULXs show strong variabilities, e.g., NGC 6946 X-1 \citep{rao10}, M82 X41.4+60 in the low state \citep{fen10}. However, their variabilities are mainly in the hard band.}
\tablenotetext{d}{possibly yes in NGC 5204 X-1 \citep{fen09}, but definitely inconsistent in NGC 1313 X-2 \citep{fen07} and IC 342 X-1 \citep{fen09}. }
\end{deluxetable}

\clearpage


\begin{thebibliography}{}

\setlength{\itemsep}{0pt}
\setlength{\parskip}{0pt}
\small

\bibitem[Begelman(2002)]{beg02}
Begelman, M.\ C.\ 2002, \apj, 568, L97

\bibitem[Belczynski et al.(2010)]{bel10}
Belczynski, K., Bulik, T., Fryer, C.\ L., Ruiter, A., Valsecchi, F., Vink, J.\ S., Hurley, J.\ R.\ 2010, \mnras, 714, 1217

\bibitem[Blustin et al.(2002)]{blu02}
{Blustin}, A.\ J., {Branduardi-Raymont}, G., {Behar}, E., {Kaastra}, J.\ S., {Kahn}, S.\ M., {Page}, M.\ J., {Sako}, M., \& {Steenbrugge}, K.\ C.\ 2002, \aap, 392, 453

\bibitem[Brandt \& Hasinger(2005)]{bra05}
Brandt, W.\ N., \& Hasinger, G.\ 2005, \araa, 43, 827

\bibitem[Carpano et al.(2007)]{car07}
Carpano, S., Pollock, A.\ M.\ T., King, A.\ R., Wilms, J., \& Ehle, M.\ 2007, \aap, 471, L55

\bibitem[Colbert \& Mushotzky(1999)]{col99}
Colbert, E.\ J.\ M., \& Mushotzky, R.\ F.\ 1999, \apj, 519, 89

\bibitem[Davidge(2006)]{dav06}
Davidge, T.\ J.\ 2006, \apj, 641, 822

\bibitem[Ebisuzaki et al.(2001)]{ebi01}
Ebisuzaki, T., Makino, J., Tsuru, T.\ G., Funato, Y., Portegies Zwart, S., Hut, P., McMillan, S., Matsushita, S., Matsumoto, H., \& Kawabe, R.\ 2001, \apj, 562, L19

\bibitem[Elvis et al.(1997)]{elv97}
Elvis, M., Fiore, F., Giommi, P., \& Padovani, P.\ 1997, \mnras, 291, L49


\bibitem[Fabbiano et al.(1992)]{fab92}
Fabbiano, G., Kim, D.-W., \& Trinchieri, G.\ 1992, \apjs, 80, 531

\bibitem[Fabbiano et al.(2003)]{fab03}
Fabbiano, G., King, A.\ R., Zezas, A., Ponman, T.\ J., Rots, A., \& Schweizer, Fran\c{c}ois 2003, \apj, 591, 843

\bibitem[Farrell et al.(2009)]{far09}
Farrell, S.\ A., Webb, N.\ A., Barret, D., Godet, O., Rodrigues, J.\ M.\ 2009, \nat, 460, 73

\bibitem[Feng \& Kaaret(2005)]{fen05}
Feng, H., \& Kaaret, P.\ 2005, \apj, 633, 1052

\bibitem[Feng \& Kaaret(2007)]{fen07}
Feng, H., \& Kaaret, P.\ 2007, \apj, 660, L113

\bibitem[Feng \& Kaaret(2009)]{fen09}
Feng, H., \& Kaaret, P.\ 2009, \apj, 696, 1712

\bibitem[Feng \& Kaaret(2010)]{fen10}
Feng, H., \& Kaaret, P.\ 2010, \apj, 712, L169

\bibitem[Ferguson et al.(1996)]{fer96}
Ferguson, A.\ M.\ N., Wyse, R.\ F.\ F., Gallagher, J.\ S., III, \& Hunter, D.\ A.\ 1996, \aj, 111, 2265

\bibitem[Gladstone et al.(2009)]{gla09}
Gladstone, J.\ C., Roberts, T.\ P., \& Done, C.\ 2009, \mnras, 397, 1836

\bibitem[Gieren et al.(2009)]{gie09}
Gieren, W., et al.\ 2009, \apj, 700, 1141

\bibitem[Grimm et al.(2003)]{gri03}
Grimm, H.\ J., Gilfanov, M., \& Sunyaev, R.\ 2003, \mnras, 339, 793

\bibitem[Gonz\'{a}lez-Mart\'{i}n et al.(2011)]{gon11}
Gonz\'{a}lez-Mart\'{i}n, O., Papadakis, I., Reig, P., \& Zezas, A.\ 2011, \aap, 526, A132

\bibitem[Isobe et al.(2008)]{iso08}
{Isobe}, N., {Kubota}, A., {Makishima}, K., {Gandhi}, P., {Griffiths}, R.\ E., {Dewangan}, G.\ C., {Itoh}, T., \& {Mizuno}, T.\ 2008, \pasj, 60, S241

\bibitem[Jin et al.(2010)]{jin10}
Jin, J., Feng, H., \& Kaaret, P.\ 2010, \apj, 716, 181

\bibitem[Kaaret et al.(2001)]{kaa01}
Kaaret, P., Prestwich, A.\ H., Zezas, A., Murray, S.\ S., Kim, D.-W., Kilgard, R.\ E., Schlegel, E.\ M., Ward, M.\ J.\ 2001, \mnras, 321, L29

\bibitem[Kajava \& Poutanen(2009)]{kaj09}
{Kajava}, J.\ J.\ E., \& {Poutanen}, J.\ 2009, \mnras, 398, 1450


\bibitem[Kalberla et al.(2005)]{kal05}
Kalberla, P.\ M.\ W., Burton, W.\ B., Hartmann, D., Arnal, E.\ M., Bajaja, E., Morras, R., \& P\"{o}ppel, W.\ G.\ L.\ 2005, \aap, 440, 775

\bibitem[Kim \& Fabbiano(2004)]{kim04}
Kim, D.\ W., \& Fabbiano, G.\ 2004, \apj, 611, 846

\bibitem[Kong et al.(2005)]{kon05}
Kong, A.\ K.\ H., \& Di Stefano, R.\ 2005, 632, L107

\bibitem[Kong et al.(2004)]{kon04}
Kong, A.\ K.\ H., Di Stefano, R., \& Yuan, F.\ 2004, 617, L49

\bibitem[Leahy et al.(1983)]{lea83}
Leahy, D.\ A., Darbro, W., Elsner, R.\ F., Weisskopf, M.\ C., Sutherland, P.\ G., Sutherland, P.\ G., Kahn, S., \& Grindlay, J.\ E.\ 1983, \apj, 266, 160

\bibitem[Lira et al.(2000)]{lir00}
Lira, P., Lawrence, A. \& Johnson, R.\ A.\ 2000, \mnras, 319, 17

\bibitem[Liu(2008)]{liu08}
Liu, J.\ 2008, \apjs, 177, 181

\bibitem[Maccacaro et al.(1982)]{mac82}
Maccacaro, T., et al.\ 1982, \apj, 253, 504

\bibitem[Maccacaro et al.(1988)]{mac88}
Maccacaro, T., Gioia, I.\ M., Wolter, A., Zamorani, G., \& Stocke, J.\ T.\ 1988, \apj, 326, 680

\bibitem[Makishima et al.(2000)]{mak00}
Makishima, K., et al.\ 2000, \apj, 535, 632

\bibitem[Margon et al.(1985)]{mar85}
Margon, B., Downes, R.\ A., \& Chanan, G.\ A.\ 1985, \apjs, 59, 23

\bibitem[McClintock \& Remillard(2006)]{mcc06}
McClintock, J.\ E., \& Remillard, R.\ A.\ 2006, in Compact Stellar X-ray sources, ed.\ W.\ H.\ G.\ Lewin \& M.\ van der Klis (Cambridge: Cambridge Univ.\ Press), 157

\bibitem[Mukai et al.(2003)]{muk03}
Mukai, K., Pence, W.\ D., Snowden, S.\ L., \& Kuntz, K.\ D.\ 2003, \apj, 582: 184

\bibitem[Mukai et al.(2005)]{muk05}
Mukai, K., Still, M., Gorbet, R.\ H.\ D., Kuntz, K.\ D., \& Barnard, R.\ 2005, \apj, 634, 1085

\bibitem[Ohsuga et al.(2005)]{ohs05}
Ohsuga, K., Mori, M., Nakamoto, T., \& Mineshige, S.\ 2005, \apj, 628, 368

\bibitem[Pence et al.(2001)]{pen01}
Pence, W.\ D., Snowden, S.\ L., Mukai, K., \& Kuntz, K.\ D.\ 2001, \apj, 561: 189

\bibitem[Poutanen et al.(2007)]{pou07}
Poutanen, J., Lipunova, G., Fabrika, S., Butkevich, A.\ G., \& Abolmasov, P.\ 2007, \mnras, 377, 1187

\bibitem[Protassov et al.(2002)]{pro02}
Protassov, R., van Dyk, D.\ A., Connors, A., Kashyap, V.\ L., \& Siemiginowska, A.\ 2002, \apj, 571, 545

\bibitem[Rao et al.(2010)]{rao10}
Rao, F., Feng, H., \& Kaaret, P.\ 2010, \apj, 722, 620

\bibitem[Read et al.(1997)]{rea97}
Read, A.\ M., Ponman, T.\ J., \& Strickland, D.\ K.\ 1997, \mnras, 286, 626

\bibitem[Reeves et al.(2003)]{ree03}
Reeves, J.\ N., O'Brien, P.\ T., \& Ward, M.\ J.\ 2003, \apj, 593, L65

\bibitem[Remillard \& McClintock(2006)]{rem06}
Remillard, R.\ A., \& McClintock, J.\ E.\ 2006, \araa, 44, 49

\bibitem[Roberts et al.(2010)]{rob10}
Roberts, T.\ P., Gladstone, J.\ C., Goulding, A.\ D., Swinbank, A.\ M., Ward, M.\ J., Goad, M.\ R., \& Levan, A.\ J.\ 2010, arXiv:1011.2155

\bibitem[Soria et al.(2007)]{sor07}
Soria, R., Baldi, A., Risaliti, G., Fabbiano, G., King, A., La Parola, V., \& Zezas, A.\ 2007, \mnras, 379, 1313

\bibitem[Soria \& Ghosh(2009)]{sor09}
Soria, R., \& Ghosh, K.\ K.\ 2009, \apj, 696, 287

\bibitem[Starrfield et al.(2004)]{sta04}
Starrfield, S., Timmes, F.\ X., Hix, W.\ R., Sion, E.\ M., Sparks, W.\ M., \& Dwyer, S.\ J.\ 2004, \apj, 612, L53

\bibitem[Stobbart et al.(2006)]{sto06}
Stobbart, A.-M., Roberts, T.\ P., \& Wilms, J.\ 2006, \mnras, 368, 397

\bibitem[Stocke et al.(1991)]{sto91}
Stocke, J.\ T., Morris, S.\ L., Gioia, I.\ M., Maccacaro, T., Schild, R., Wolter, A., Fleming, T.\ A., \& Henry, J.\ P.\ 1991, \apjs, 76, 813

\bibitem[Swartz et al.(2002)]{swa02}
Swartz, D.\ A., Ghosh, K.\ K., Suleimanov, V., Tennant, A.\ F., \& Wu, K.\ 2002, \apj, 574, 382

\bibitem[Volonteri(2010)]{vol10}
Volonteri, M.\ 2010, Astron.\ Astrophys.\ Rev., 18, 279

\bibitem[Watarai et al.(2001)]{wat01}
Watarai, K.-Y., Mizuno, T., \& Mineshige, S.\ 2001, \apj, 549, L77

\bibitem[Winter et al.(2006)]{win06}
Winter, L.\ M., Mushotzky, R.\ F., \& Reynolds, C.\ S.\ 2006, \apj, 649, 730

\bibitem[Zang et al.(1997)]{zan97}
Zang, Z., Warwick, R.\ S., \& Meurs, E.\ J.\ A.\ 1997, Irish Astr. J., 24, 45

\bibitem[Zhou et al.(2010)]{zho10}
Zhou, X.-L., Zhang, S.-N., Wang, D.-X., \& Zhu, L.\ 2010, \apj, 710, 16

\end{thebibliography}
\end{document}